\documentclass{ifacconf}

\usepackage{graphicx}      % include this line if your document contains figures
\usepackage{natbib}        % required for bibliography
\usepackage{amsmath,amssymb,amsfonts}
\usepackage{mathrsfs}
\usepackage{enumitem}
\usepackage{nicefrac}
\usepackage{subfig}
\usepackage{multirow}

\newtheorem{remark}{Remark}
\begin{document}

\onecolumn
\begin{center}
Accepted for publication at the IFAC World Congress 2023

\copyright 2023 the authors. This work has been accepted to IFAC for publication under a Creative Commons Licence CC-BY-NC-ND

\end{center}

\begin{frontmatter}

\title{Model-Free Control Design Procedure Applied to Lateral Vehicle Control} %\thanksref{footnoteinfo}

%\thanks[footnoteinfo]{This work is partially supported by the joint CNRS (France)-CSIC (Spain) PICS Program through the project UbiMFC (CoopIntEer 203 073)}%

\author[First]{Marcos Moreno-Gonzalez}
\author[First]{Jorge Villagra}

\address[First]{Centre for Automation and Robotics (CAR), CSIC - Universidad Politécnica de Madrid, ctra. de Campo Real, km 0,200, 28500 Arganda del Rey, Spain, (e-mail: \{marcos.moreno, jorge.villagra\}@csic.es).}

\begin{abstract}                % Abstract of not more than 250 words.
Model-Free Control has proven its performance in a wide variety of systems. Although its adequate tuning can be achieved using the knowledge of the system and optimization-based approaches, there is not yet a systematic design procedure for this kind of control scheme. In this paper, a non-iterative Three Term Controller tuning procedure is adapted and extended to fit Model-free controllers' structure. This procedure is successfully applied to design the lateral control of an automated car with realistic performance requirements.
\end{abstract}


\end{frontmatter}
%===============================================================================

\section{Introduction}

Model-Free Control (MFC) \citep{fliess.ijc13} has gained attention over the last years as a control technique able to successfully regulate complex systems that are time-varying or non-linear, e.g., \cite{fliess2021alternative}, \cite{villagra.ijvas09} and \cite{villagra2020model} are some examples that show this potential. Variations of the original control structure can also be found in the literature, such as \cite{wang2020alpha} and \cite{moreno2022speed}. However, although some recent works propose different solutions \citep{Yahagi2022,hegedHus2022design}, the design of these \textit{intelligent} controllers is still an open problem. %In \cite{d2010mathematical}, some sort of equivalence between PID and MFC controllers is found under certain circumstances. Although it is not its purpose, this result can be exploited to tune MFC controllers, but it may have limitations, as discussed below.

In this paper, a new control design procedure for regulators under the Model-Free Control paradigm is presented. The proposed algorithm consists on the adaptation of the discrete Three Term Controller design procedure collated in \cite{bhattacharyya2018linear}. It is based on the root-counting and phase unwrapping formulas for discrete-time systems detailed in \cite{keel2002root}, and allows the designer to obtain the stability region of the controller given an approximate model of the system.

To evaluate the potential of the MFC design procedure, the lateral control of an automated car is studied, applying appropriate restrictions for the control design and using a realistic vehicle simulator.

The rest of the paper is structured as follows. Section~\ref{PID_des} presents the Three Term Controller design procedure which is later adapted to MFC. A brief introduction to Model-Free Control is presented in Section~\ref{MFC_princ}. Section~\ref{MFC_design} details the proposed adaptation from PID design to Model-Free Control design. An illustrative example and the results from simulation are presented in Section~\ref{Results}. Finally, concluding remarks and references can be found in the last section.

\section{Three Term Controller Design Procedure}\label{PID_des}

This section briefly explains the Three Term Controller design procedure that provides a stabilizing set of control parameters for a given plant model that will be used in this work. The design procedure is taken from \cite{bhattacharyya2018linear} and the references therein, where a complete description of the algorithm can be found.

Consider a discrete-time system $G(z) = N(z)\big/D(z)$ where $N(z)$ and $D(z)$ are real polynomials, being $deg[D(z)]=n_1$ and $deg[N(z)]=l_1\leq n_1$. The plant is controlled by a Two Term Controller (which is similar to a PI controller):
\begin{equation}
C_{PI}(z)=K_1\frac{z-K_2}{z-1}
\label{eq_PIctrl}
\end{equation}
\noindent or a Three Term Controller (similar to a PID controller):
\begin{equation}
C_{PID}(z)=\frac{K_2 z^2 + K_1 z +K_0}{z(z-1)}
\label{eq_PIDctrl}
\end{equation}
The characteristic equation of the closed-loop system $\delta (z)$ can be expressed for each controller as follows:
\begin{align*}
\delta_{PI} (z) & = (z-1)\cdot D(z)+K_1 (z-K_2)\cdot N(z) \\
\delta_{PID} (z) & = z(z-1)\cdot D(z)+(K_2 z^2 + K_1 z +K_0)\cdot N(z) \label{eq_delta}
\end{align*}
The parameters that stabilize the closed-loop system, i.e., the stability region, can be found following these steps:

\begin{enumerate}[label=\arabic*)]
\item Define $N(z)$ and $D(z)$ and find their Tchebychev representation:
\begin{equation*}
D(z) = \sum_{k=0}^{n_1} a_k \cdot z^k; \;\; N(z) = \sum_{k=0}^{l_1} b_k \cdot z^k
\end{equation*}
\begin{equation*}
\begin{split}
N(z)\Big|_{z=e^{j\theta}} \!\!\!\!\! & = N(e^{j\theta})\Big|_{\cos{\theta} =u} \!\!\!\!\! = R_N(u) + j\sqrt{1-u^2}T_N(u)\\ 
D(z)\Big|_{z=e^{j\theta}} \!\!\!\!\! & = D(e^{j\theta})\Big|_{\cos{\theta} =u} \!\!\!\!\! = R_D(u) + j\sqrt{1-u^2}T_D(u)
\end{split}
\label{eq_Tcheby}
\end{equation*}
\begin{equation*}
R_{D}(u) = \sum_{k=1}^{n_1} a_k \cdot c_k(u) + a_0; \;\; T_{D}(u) = \sum_{k=1}^{n_1} a_k \cdot s_k(u)
\end{equation*}
\begin{equation*}
R_{N}(u) = \sum_{k=1}^{l_1} b_k \cdot c_k(u) + b_0; \;\; T_{N}(u) = \sum_{k=1}^{l_1} b_k \cdot s_k(u)
\end{equation*}
where $c_k$ and $s_k$ are defined in Table \ref{table:Tcheby}.
\begin{table}[h]
    \centering
    \begin{tabular}{c|c c}
        $k$ & $c_k(u)$ & $s_k(u)$ \\ \hline 
        1 & $-u$ & $1$ \\
        2 & $2u^2 -1$ & $-2u$ \\
        3 & $-4u^3 +3u$ & $4u^2 -1$ \\
        4 & $8u^4 -8u^2 +1$ & $-8u^3 +4u$ \\
        5 & $-16 u^5 +20u^3 -5u$ & $16u^4 -12u^2 +1$ \\
        \vdots & \vdots & \vdots
    \end{tabular}
    \captionsetup{width=1.1\linewidth}
    \caption{Generalized Tchebychev polynomials}
    \label{table:Tcheby}
\end{table}
\item Define $\nu (z)$ for each controller from the corresponding characteristic equation and find its Tchebychev representation:
\begin{equation*}
\begin{split}
\nu_{PI} (z) & = \delta(z) N(z^{-1}) \\
        & = R(u, K_{1, 2}) + j\sqrt{1-u^2} \cdot T(u, K_1)
\end{split}
\label{eq_nu1}
\end{equation*}
\begin{equation}
\begin{split}
R(u, K_{1, 2}) = & -(u + 1)P_1(u) -(1 - u^2)P_2(u) \\
                 &- K_1 (u + K_2) P_3(u) \\
T(u, K_1) = & K_1 P_3(u) + P_1(u) - (u + 1)P_2(u)
\end{split}
\label{eq_RT1}
\end{equation}
\begin{equation*}
\begin{split}
\nu_{PID} (z) & = z^{-1}\delta(z) N(z^{-1}) \\
        & = R(u, K_{1, 2, 3}) + j\sqrt{1-u^2} \cdot T(u, K_3)
\end{split}
\label{eq_nu2}
\end{equation*}
\begin{equation}
\begin{split}
R(u, K_{1, 2, 3}) = \, & -(u + 1)P_1(u) -(1 - u^2)P_2(u) \\
                 &- \left[(2K_2 - K_3)u - K_1)\right] P_3(u) \\
T(u, K_3) = \,& K_3 P_3(u) + P_1(u) - (u + 1)P_2(u)
\end{split}
\label{eq_RT2}
\end{equation}
where $K_3$ is defined as $K_3 \triangleq K_2-K_0$. 

\noindent Note that $T(\cdot)$ only depends on one control parameter, either $K_1$ or $K_3$.

\noindent Obtain $P_1(u)$, $P_2(u)$ and $P_3(u)$:
\begin{equation*}
\begin{split}
P_1 (u) & = R_D (u) R_N (u) + (1-u^2) T_D (u) T_N (u) \\ 
P_2 (u) & = R_N (u) T_D (u) - R_D (u) T_N (u) \\
P_3 (u) & = R_N ^2 (u) + (1-u^2) T_N ^2 (u)
\end{split}
\label{eq_pes}
\end{equation*}
\item Calculate the signature of $\nu (z)$ as $\sigma$:
\begin{align*}
\sigma_{PI} & = i_\delta + i_{N_r} - l_1 \\
\sigma_{PID} & = i_\delta + i_{N_r} - l_1 - 1
\label{eq_signature}
\end{align*}
where $i_\delta$ and $i_{N_r}$ are the number of roots inside the unit circle of $\delta(z)$ and $N_r (z)$ respectively, and $N_r (z)$ is defined as $N_r (z) = N(z^{-1}) \cdot z^{l_1}$. Note that all the roots of $\delta(z)$ must be inside the unit circle to ensure stability ($i_\delta = deg[\delta (z)]$).

\item Obtain the set of $K_1$ or $K_3$ as appropriate ($K_{1(3)}$) such that $T(u,K_{1(3)})$ from \eqref{eq_RT1} or \eqref{eq_RT2} have at least $\sigma - 1$ real different zeros of odd multiplicity when $u \in(-1,1)$. For this purpose, root locus can be applied as $T(\cdot)$ only depends on one parameter.

\item If the set of $K_{1(3)}$ is empty, then there is no stabilizing set for the system with the type of controller chosen.

\end{enumerate}
For each $K_{1(3)}$:

\begin{enumerate}[label=\arabic*)]
\setcounter{enumi}{5}
\item Find the $k$ real distinct zeros of odd multiplicity $t_i$ of $T(u)$ (\eqref{eq_RT1} or \eqref{eq_RT2}) for $u \in (-1,1)$ and arrange them as $-1 < t_1 < t_2 < \dots < t_k < +1$.

\item Build the set of sequences (known as \textit{strings}) $A_k=\{\{i_0,i_1,...,i_k,i_{k+1}\}\}$ where each $i_j$ has the value $-1$ or $1$, $k$ is the number of real distinct zeros of odd multiplicity of $T(u)$ from the previous step and $A_k$ covers all the possible combinations.

\item Determine the admissible string set $I\in A_k$ in which each admissible string satisfies: 
\begin{equation*}
\sigma=\frac{sgn\!\left[T^{(p)} (-1)\right]}{2} \!\!\left(\!\!i_0 + 2\!\sum_{j=1}^{k} (-1)^j i_j +\!(-1)^{k+1} i_{k+1}\!\!\right)
\label{eq_strings}
\end{equation*}
where $p$ is the number of zeros of $T(u)$ in $u=-1$.

\end{enumerate}
For each admissible string $I_i$:

\begin{enumerate}[label=\arabic*)]
\setcounter{enumi}{8}
\item Determine the values of $K_2$ (and $K_1$ for PID) that simultaneously satisfy the inequalities:
\begin{equation*}
R(t_j) \cdot i_t >0 \;\;\; \forall t=0, 1, \dots , k, k\!+\!1
\label{eq_inequalities}
\end{equation*}
\item Only in the PID case, obtain $K_0$ from $K_3=K_2-K_0$.
\end{enumerate}

\begin{remark}
The stabilizing set obtained with this procedure contains all and only the control parameters that make the closed-loop system stable. However, concrete values of $K_{1(3)}$ are needed for steps 6 to 10, so a discretization has to be made which can cause a loss of information (see Fig. \ref{fig:stabilizingset} for an illustrative example).
\end{remark}

\section{Model-Free Control principles}\label{MFC_princ}

In \cite{fliess.ijc13}, it is shown that the dynamics of a system can be replaced by an ultra-local model 
\begin{equation}
y^{(n)} = F + \alpha \cdot u %\overset{n}{\dot{y}} sería la forma de poner la derivada enésima coherente con la notación de punto, pero es menos conocida
\label{eq_ultralocal}
\end{equation}
in which the input $u$ and the nth time derivative of the output $y$ are linearly related by a constant design parameter $\alpha$ and the relationship is fitted by $F$, a variable that absorbs system disturbances and model errors. Note that $n$ defines the order of the ultra-local model.

The control loop is typically closed by a so-called \textit{intelligent PID} controller, iPID controller (usually iP or iPD):
\begin{equation}
u = \frac{1}{\alpha}  \cdot \left(-F + y_r ^{(n)} + K_p \, e + K_i \int{e} + K_d \, \dot{e}\right) 
\label{eq2}
\end{equation}
where $u$ is the control action, suffix $r$ stands for reference, $e$ is the error and $K_p$, $K_i$ and $K_d$ are the PID control parameters. $F$ is estimated in real time using an estimator $\hat{F}$; the simplest one assumes it to be constant between consecutive instants and can be estimated from \eqref{eq_ultralocal} as: 
\begin{equation}
\hat{F} (t_k) = \hat{y} ^{(n)} (t_k) - \alpha \cdot u (t_{k - 1}) 
\label{eq3}
\end{equation}
where $t_k$ is the current instant and $\hat{y} ^{(n)}$ is the filtered nth time derivative of $y$. More elaborated estimators for discrete-time settings can be found in \cite{sanyal2022discrete}.

\begin{remark}
Note that the error dynamics derived from (\ref{eq_ultralocal}) and (\ref{eq2}) can be expressed as $e^{(n)}+K_d \dot{e}+K_p e+K_i \int{e}=\hat{F}-F$. If the estimation of $F$ is good enough ($\hat{F} \approx F$), then the system dynamics can be made asymptotically stable through an appropriate choice of the control parameters and order of the ultra-local model.
\end{remark}
\section{MFC design procedure}\label{MFC_design}
\subsection{Justification}\label{section:justif}
The developed design procedure is based on the relationship between Model-free controllers and the Three Term Controllers from Section \ref{PID_des}. 

Although MFC has advantages over PID control, both in terms of performance and robustness (cf. \cite{li2022revisit}), a mathematical equivalence between PI(D) and iP(D) control structures can be demonstrated under certain assumptions (e.g., assuming that the derivative of the output of the system is not filtered, \cite{d2010mathematical}). However, real measurements are usually filtered before derivation to prevent the control action \eqref{eq2} from chattering (as it depends on the derivative of the system output).% and can be seen in Fig..

%Although it is not common, it can reach the extreme when a PID controller with an aggressive configuration transforms into an unstable iPD configuration. %Table \ref{tab:iPD_clasic} gives an example where an unstable iPD is obtained with this transformation for the vehicle's lateral dynamics in \eqref{eq_latdyn} (which will be used later to test the proposed MFC design algorithm). Note that the PID tested here may become unstable in a realistic application due to the saturation of the actuators.

%\begin{table}
%\centering
%\caption{Control parameter equivalence between PID and iPD of second order (sample time $T_s = 0.05 \,s$)}
%\label{tab:iPD_clasic}
%\begin{tabular}{|c|c|c|c|} 
%\hline
%\multicolumn{2}{|c|}{PID Controller} & \multicolumn{2}{c|}{iPD$_2$ Controller }  \\ \hline
%$K_P$   & 2.14                  & $\alpha = 1/T_s K_D$  & 31.69                          \\ \hline
%$K_I$   & 5.48                  & $K_p = K_I T_s \alpha$     & 8.683                          \\ \hline
%$K_D$   & 0.674                 & $K_d = K_P T_s \alpha$     & 3.393                          \\\hline
%\end{tabular}
%\end{table}

%\begin{figure}
%    \centering
%    \subfloat[PID controller]{
%    \includegraphics[width=0.75\linewidth]{figures/PID_example.pdf}}
%    
%    \subfloat[Equivalent iPD controller]{
%    \includegraphics[width=0.75\linewidth]{figures/iPD_clasico.pdf}}
%    
%    \caption{Example of PID - iPD classical transformation}
%    \label{fig:iPD_clasic}
%\end{figure}
%
In the following sections, the equivalence between a first (second) order iPD controller and a Two (Three) Term Controller (corresponding to PI and PID controllers) is shown considering a filtered derivative of the system output, as requested in \eqref{eq3}. This derivative can be expressed as the following operator:
\begin{equation}
D (z) = \frac{1}{T_s} \frac{z-1}{C z+1-C}
\label{eq_filter}
\end{equation}
\noindent where $T_s$ is the sampling time and $C$ is a filtering parameter, which is characterized according to the measured signal-noise ratio.

Notice that only iPD controllers are considered in this work, as they are the most usual \citep{fliess2021alternative}.

\subsection{First Order iPD and PI}\label{section:first_order}
From \eqref{eq2}, \eqref{eq3} and \eqref{eq_filter}, and considering a first order ultra-local model, the discrete transfer function $C_{iPD_1}(z)$ of the iPD controller, that relates the control action $U(z)$ to the error $E(z)$, can be obtained:
\begin{equation}
C_{iPD_1} (z) \!= \!\tfrac{U_{iPD_1} (z)}{E(z)} \!=\! \frac{\frac{K_p T_s C + K_d + 1}{\alpha T_s} z^2 \!- \frac{K_p T_s (C-1) + K_d + 1}{\alpha T_s} z}{(z-1) \cdot (C z+1-C)}
\label{eq_iPD1}
\end{equation}
\noindent which in turn yields:
\begin{equation*}
C_{iPD_1} (z) = K_1 \frac{z-K_2}{z-1} \cdot \frac{z}{C z+1-C}
\label{eq_iPD1_PI}
\end{equation*}
\noindent where 
\begin{equation}
K_1 = \frac{K_p T_s C + K_d + 1}{\alpha T_s}; \; \; K_2 = \frac{K_p T_s (C-1) + K_d + 1}{\alpha T_s}
\label{eq_K1K2}
\end{equation}
Comparing the direct loops $C_{PI} (z) \cdot G(z)$ and $C_{iPD_1} (z) \cdot G(z)$, it is obtained that controlling a system $G(z)$ with the iPD$_1$ controller from \eqref{eq_iPD1} is the same as controlling $G(z) \cdot \dfrac{z}{C z+1-C}$ with the Two Term (PI) controller from~\eqref{eq_PIctrl}. Additionally, an inverse relationship can be obtained from~\eqref{eq_K1K2}:
\begin{equation}
\frac{K_p}{\alpha} = K_1 \cdot (1-K_2); \; \; \frac{K_d+1}{\alpha} = T_s K_1 \cdot (1-C+K_2 C)
\label{eq_KpKd}
\end{equation}
\begin{remark}
Note that \eqref{eq_KpKd} is a non-linear equation in which one of the iPD control parameters is undetermined.
\end{remark}
\subsection{Second Order iPD and PID}\label{section:second_order}
Alternatively, if a second order ultra-local model is considered, the transfer function of the controller can be expressed as follows:
\begin{equation}
C_{iPD_2} (z) = z \tfrac{K_p T_s^2 (C z+1-C)^2 + K_d T_s(z-1)(C z+1-C) + (z-1)^2}{\alpha T_s^2 (z-1)(C z+1-C)^2}
\label{eq_iPD2}
\end{equation}
\noindent which yields:
\begin{equation*}
C_{iPD_2} (z) = \frac{K_2 z^2 + K_1 z + K_0}{z(z-1)} \cdot \frac{z^2}{(C z+1-C)^2}
\label{eq_iPD2_PID}
\end{equation*}
\indent where 
\begin{equation}
\begin{split}
K_2 & = \frac{K_p T_s^2 C^2 + K_d T_s C + 1}{\alpha T_s^2} \\ 
K_1 & = \frac{2 K_p T_s^2 C(1-C) + K_d T_s (1-2C) - 2}{\alpha T_s^2} \\ 
K_0 & = \frac{K_p T_s^2 (C-1)^2 + K_d T_s (C-1) + 1}{\alpha T_s^2}
\end{split}
\label{eq_K2K1K0}
\end{equation}
As in the previous section, comparing the direct loops when Three Term (PID) and iPD$_2$ controllers are applied to the plant, it is obtained that controlling a system $G(z)$ with the iPD$_2$ from \eqref{eq_iPD2} is the same as controlling $G(z) \cdot \dfrac{z^2}{(C z+1-C)^2}$ with the PID from \eqref{eq_PIDctrl}. Additionally, there are two inverse relationships among the control parameters: the semi-linear \eqref{eq_KpKdalfa1} and the non-linear \eqref{eq_KpKdalfa2}.
\begin{equation}
\begin{bmatrix}
\nicefrac{K_p}{\alpha}\\
\nicefrac{K_d}{\alpha}\\
\nicefrac{1}{\alpha}
\end{bmatrix}
=
\begin{bmatrix}
1 & 1 & 1\\
2T_s (1-C) & T_s (1-2C) & -2T_s C\\
T_s^2 (C-1)^2 & T_s^2 (C^2-C) & T_s^2 C^2
\end{bmatrix}
\begin{bmatrix}
K_2\\
K_1\\
K_0
\end{bmatrix}
\label{eq_KpKdalfa1}
\end{equation}
\begin{equation}
\begin{bmatrix}
T_s^2 C^2 & T_s C & -T_s^2 K_2\\
T_s^2 2C(1-C) & T_s(1-2C) & -T_s^2 K_1\\
T_s^2 (C-1)^2 & T_s (C-1) & -T_s^2 K_0
\end{bmatrix}
\begin{bmatrix}
K_p\\
K_d\\
\alpha
\end{bmatrix}
=
\begin{bmatrix}
-1\\
2\\
-1
\end{bmatrix}
\label{eq_KpKdalfa2}
\end{equation}

\section{Results}\label{Results}
In this section, the Model-free controller design procedure presented in previous sections is applied to tune the lateral controller of an autonomous vehicle.
\subsection{Vehicle simplified model}\label{section:example}
Vehicle's lateral dynamics are linearly modeled as proposed in \cite{rajamani2011vehicle}: 
\begin{equation}
\begin{split}
\begin{bmatrix}
\dot{e}_y\\
\ddot{e}_y\\
\dot{e}_\psi\\
\ddot{e}_\psi
\end{bmatrix}=&
\begin{bmatrix}
0 & 1 & 0 & 0\\
0 & -\tfrac{2C_f + 2C_r}{m v_x} & \tfrac{2C_f + 2C_r}{m} & \tfrac{-2C_f l_f + 2C_r l_r}{m v_x}\\
0 & 0 & 0 & 1\\
0 & \tfrac{2C_r l_r - 2C_f l_f}{I_z v_x} & \tfrac{2C_f l_f - 2C_r l_r}{I_z} & \tfrac{-2C_f l_f^2 - 2C_r l_r^2}{I_z v_x}
\end{bmatrix}
\begin{bmatrix}
e_y\\
\dot{e}_y\\
e_\psi\\
\dot{e}_\psi
\end{bmatrix}\\
& +
\begin{bmatrix}
0\\
\tfrac{2C_f}{m}\\
0\\
\tfrac{2C_f l_f}{I_z}
\end{bmatrix}
\delta
+
\begin{bmatrix}
0\\
-\tfrac{2C_f l_f - 2C_r l_r}{m v_x}-v_x\\
0\\
-\tfrac{2C_f l_f^2 + 2C_r l_r^2}{I_z v_x}
\end{bmatrix}
\dot{\psi}_{des}
\end{split}
\label{eq_latdyn}
\end{equation}
where $e_y$ and $e_\psi$ are the lateral and angular errors, $m$ is the vehicle mass, $v_x$ is the longitudinal speed, $C_r$ and $C_f$ are the cornering stiffness of the rear and front wheels, $l_r$ and $l_f$ is the distance between the center of gravity (CoG) and the rear and front axle, $I_z$ is the yaw inertia, $\delta$ is the steering angle and $\dot{\psi}_{des}$ is the desired yaw rate. Although real vehicles are substantially more complex and have strong non-linearities and varying parameters, this model is commonly used in lateral control design \citep{zainal2017yaw, jiang2018lateral}.

The state-space model is converted to a transfer function with the vehicle parameters in Table \ref{tab:vehicle_params}. It takes the lateral error and the steering angle as the output and input of the system, respectively, and the term related to the desired yaw rate is considered as a disturbance. The transfer function is discretized with a zero-order hold and a sample time $T_s = 0.05\,s$. With the discrete-time transfer function, the procedure detailed in sections \ref{PID_des} and \ref{section:second_order} can be applied.

\begin{table}[htbp]
\centering
\captionsetup{width=1.1\linewidth}
\caption{Vehicle parameters}
\begin{tabular}{|c|c|c|c|c|c|c|c|} 
\hline
 Parameter  & $m$    & $v_x$ & $I_z$ & $C_f$ & $C_r$ & $l_f$ & $l_r$ \\ \hline
 Value (IS) & \!1372\!   & 9.72    & \!1990\!  & \!37022.5\! & \!35900\! & 0.98  & 1.48 \\ \hline
\end{tabular}
\label{tab:vehicle_params}
\end{table}

\subsection{Vehicle extended model}\label{section:simulator}

The model of the vehicle used in simulation tests is the same used in \citep{moreno2022speed}. It was designed to mimic an experimental platform (described in \cite{artunedo2019decision}) with a high degree of fidelity. For this purpose, a dynamic model with 14 degrees of freedom was considered (6 for the vehicle body motion: longitudinal, lateral, vertical, roll, pitch, and yaw; and 8 for the wheels: vertical motion and spin of each wheel).

The power-train modeling comprises three elements: (i) the engine, whose torque map has been identified from measurements taken in the experimental platform; (ii) the gearbox, that reproduces the same drive ratios and gear shifting logic of the actual vehicle; (iii) the resistance torques coming from braking system, longitudinal wind forces and gravitational forces.
The tire behaviour was reproduced with the Pacejka tire model~\citep{Pacejka1992}.

Special attention was paid to the steering actuator modeling. An external actuation system was added in addition to the vehicle's electric power assisted steering system, which is modeled inspired on~\cite{7886365}. The main parameters of the modeled actuator, such as inertia or backlash, have been measured or identified from extensive field tests.
Moreover, the small noise coming from the localization system of the experimental platform has been also characterized and included in the simulation model.

\subsection{Simulation results}

The derivative filter is experimentally designed to reduce the noise of the measured lateral error, resulting in $C=4$.

\begin{figure}
    \centering
    \includegraphics[width=0.9\linewidth]{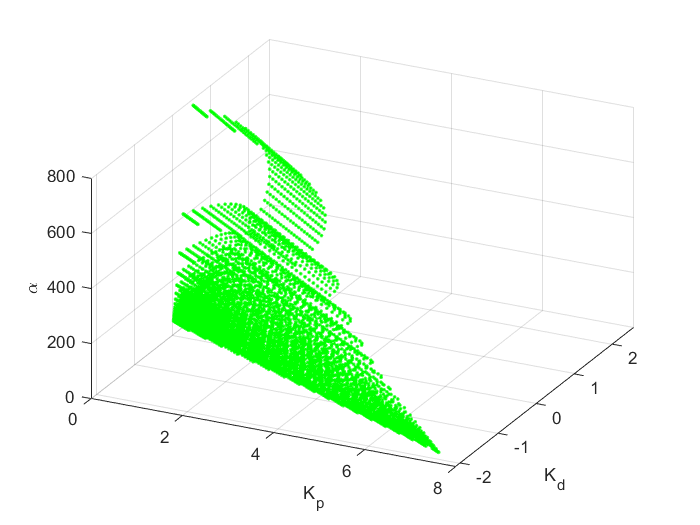}
    \caption{iPD$_2$ stabilizing set obtained with \eqref{eq_KpKdalfa2}}
    \label{fig:stabilizingset}
\end{figure}

Applying the Model-free controller design procedure presented in Section \ref{section:second_order}, i.e., obtaining the Three Term Controller stabilizing set for the equivalent system $G(z) \cdot \frac{z^2}{(C z+1-C)^2}$ and then applying transformations \eqref{eq_KpKdalfa1} or \eqref{eq_KpKdalfa2}, the iPD$_2$ stabilizing set for the plant model $G(z)$ is obtained, where $G(z)$ is the discrete-time representation of \eqref{eq_latdyn} obtained as described in Section \ref{section:example}. The stabilizing set obtained is represented in Figure \ref{fig:stabilizingset}. Note that the non-linear transformation \eqref{eq_KpKdalfa2} is applied because it retrieves the real control parameters ($\alpha, K_p, K_d$), although the resolution of the original PID stabilizing set affects the iPD parameters.

To show that the transformation keeps the stability features assured by the initial design method, several tests are made. First, the iPD$_2$ stabilizing set from Fig.~\ref{fig:stabilizingset} is checked to be in fact stable through a step response simulation. It is thus shown that the transformation does not worsen the stability characteristics of the closed-loop system.

\begin{figure}[ht]
    \centering
    \subfloat[Frequency-response restricted iPD$_2$ subset]{
    \includegraphics[width=0.95\linewidth]{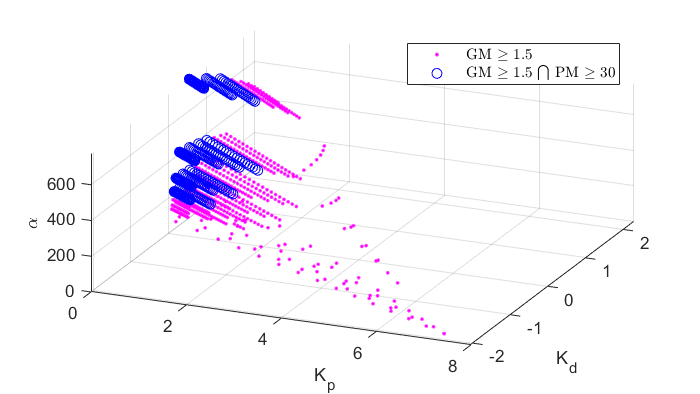}\label{restricted_set:A}}
    
    \subfloat[Time-response restricted iPD$_2$ subset]{
    \includegraphics[width=0.95\linewidth]{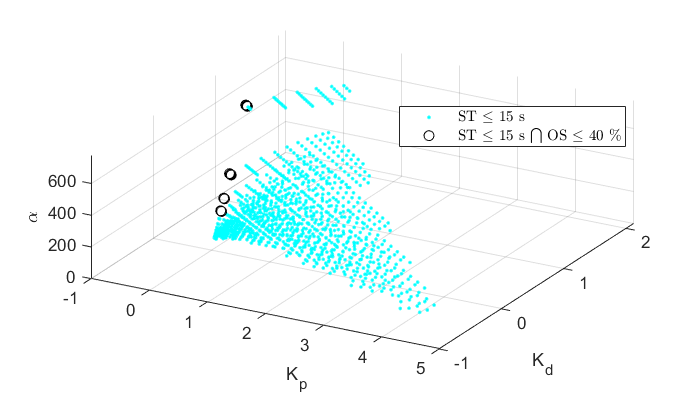}\label{restricted_set:B}}
    
    \caption{Restricted iPD$_2$ subsets}
    \label{restricted_set}
\end{figure}

\begin{figure}[ht]
\centering
\includegraphics[width=0.9\linewidth]{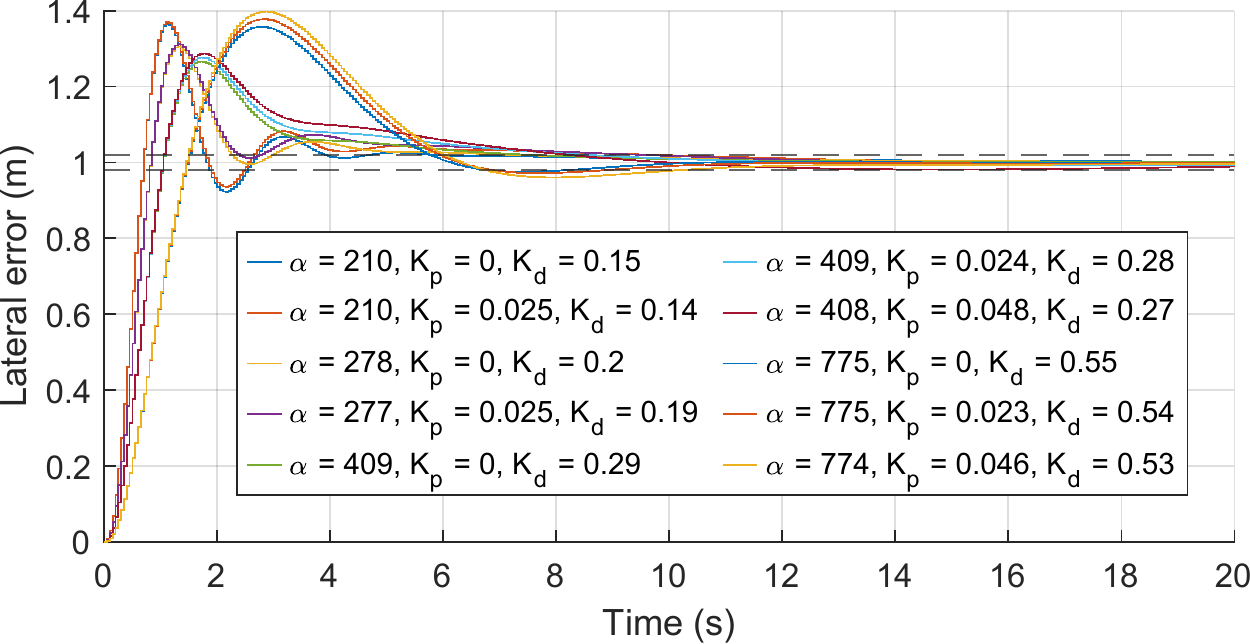}
\caption{Step response of restricted iPD$_2$ configurations}
\label{fig:step}
\end{figure}

Besides, two subsets of PID configurations that meet more demanding frequency and time-response restrictions (subset 1: gain margin, GM~$\geq 1.5$, phase margin, PM~$\geq 30\,deg$; subset 2: overshoot, OS~$\leq 40\,\%$, settling time, ST~$(2\,\%) \leq 15\, s$) are obtained and transformed into iPD$_2$ parameters. The restricted subsets are represented in blue in Fig. \ref{restricted_set:A} and in black in Fig. \ref{restricted_set:B}, respectively. The frequency-response restricted iPD$_2$ subset is checked to meet the restrictions given. The step response of the time-response restricted iPD subset is simulated in Fig. \ref{fig:step}. It is thus shown that the transformation lets the second-order iPD controllers keep more strict features of the equivalent Three Term Controllers.

The applicability of the design method is tested on the extended vehicle model detailed in Section \ref{section:simulator}. Setting different frequency and time-response restrictions, different restricted subsets are obtained; e.g., the subsets that Fig. \ref{restricted_set} show. Among those restricted subsets, four different iPD$_2$ configurations are chosen, which are gathered in Table \ref{tab:iPD_params}. Note that the time-response specifications are obtained from the step response.

\begin{table}[ht]
\centering
\captionsetup{width=1.1\linewidth}
\caption{iPD control parameters tested}
\begin{tabular}{|c|c|c|c|c|c|c|c|} 
\hline
\begin{tabular}[c]{@{}c@{}}\!\!iPD\!\!\\\!\!Ctrl.\!\!\end{tabular} & $K_p$   & $K_d$   & $\alpha$ & \!\!OS \!(\%)\!\! & \!\!ST \!(s)\!\! & \!\!GM \!(dB)\!\! & \!\!PM \!(º)\!\!    \\ \hline
1   & \!\!0.00093\!\! & \!0.043\! & \!315.7\!    & 10-20  & 0-10  & -     & - \\ \hline
2   & \!\!0.09078\!\! & \!0.167\!  & \!161.9\!    & 40-50  & 10-20 & -     & - \\ \hline
3   & 0.0     & \!0.301\!  & \!116.1\!    & -      & -     & 0-5   & 0-10   \\ \hline
4   & 0.0     & \!0.649\!  & \!792.6\!    & -      & -     & 20-30 & 50-60  \\ \hline
\end{tabular}
\label{tab:iPD_params}
\end{table}

Real vehicles must not exceed certain dynamic constraints -such as in lateral acceleration (\cite{villagra2023interplay}- in order to assure stability, so a step response test is unfeasible. Instead, a smooth trajectory is defined to evaluate the controller performance. 

\begin{figure}
\centering
\includegraphics[width=0.85\linewidth]{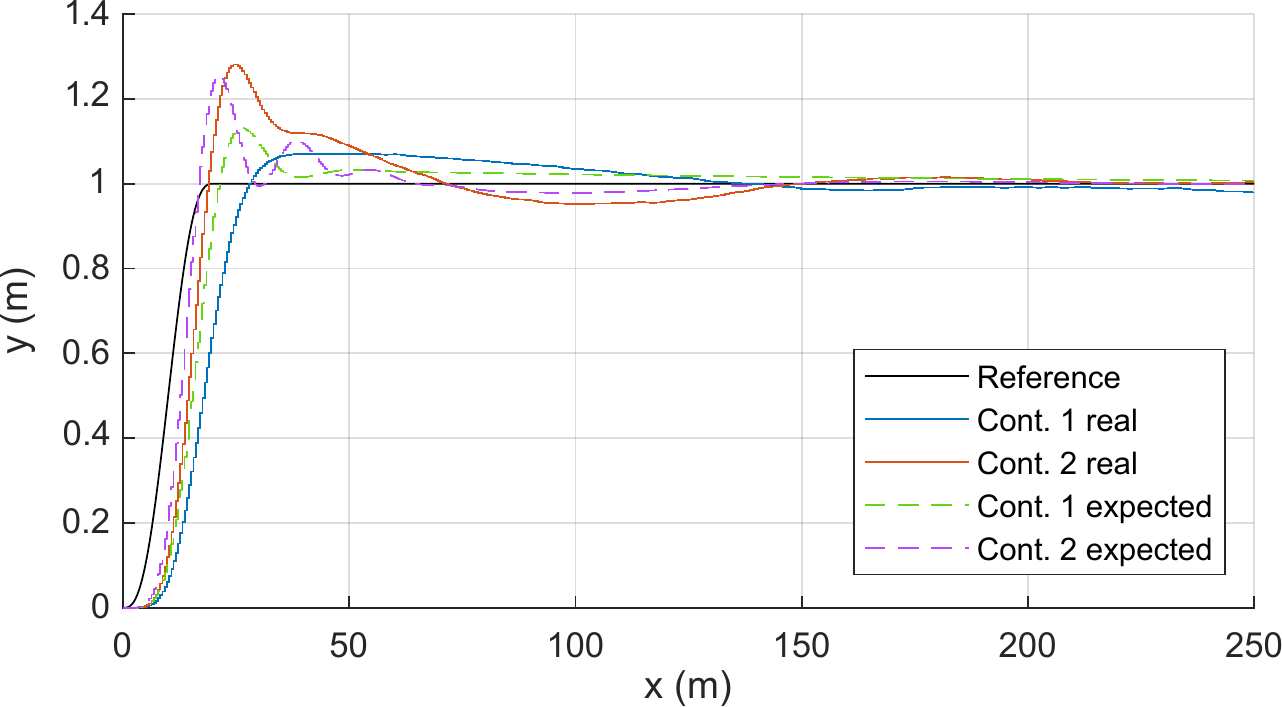}
\caption{Step responses of controllers 1 and 2}
\label{fig:smoothstep}
\end{figure}

Controllers 1 and 2 are tested on the extended model using a smoothed step trajectory, setting the initial longitudinal speed of the vehicle at $35\, km/h$. Table \ref{tab:time_resp_1_2} gathers the time-response parameters (obtained with the simplified model \eqref{eq_latdyn}) and with the extended model). Figure \ref{fig:smoothstep} shows the response of both models with controllers 1 and 2. Note that the differences between the expected and real settling time can be caused by the steering actuator dynamics, that are neglected in the simplified vehicle model.

\begin{table}
\centering
\captionsetup{width=1.1\linewidth}
\caption{Time-response restrictions}
\begin{tabular}{|c|c|c|c|c|} 
\hline
\multirow{2}{*}{\begin{tabular}[c]{@{}c@{}}iPD\\Controller\end{tabular}} & \multicolumn{2}{c|}{OS ($\%$)} & \multicolumn{2}{c|}{ST (5\,\%) (s)}  \\ 
\cline{2-5}
                                             & Expected & Real          & Expected & Real          \\ \hline
1                                            & 13.15      & 7.06        & 3.35       & 8.55  \\ \hline
2                                            & 25.35      & 28.12       & 4.45       & 5.95  \\ \hline
\end{tabular}
\label{tab:time_resp_1_2}
\end{table}
On the other hand, evaluation of the gain and phase margins on the extended model is not straightforward as it is highly non-linear. Frequency-response margins are related to the absolute stability of the system, therefore it is tested instead by simulating the behavior of controllers 3 and 4 on test trajectories that impose different dynamic constraints on the vehicle, as can be seen in Table \ref{tab:trajectories}. Note that $T_2$ is much more demanding than $T_1$ in maximum speed and acceleration constraints. This test trajectories are obtained by applying the speed planning method proposed in \cite{Artunedo2021}.

\begin{table}[htpb]
\centering
\captionsetup{width=1.1\linewidth}
\caption{Dynamic constraints by trajectory}
\label{tab:my-table}
\begin{tabular}{|l|l|l|}
\hline
Trajectory                       & $T_1$  & $T_2$ \\ \hline
Maximum speed ($km/h$)               & 35  & 100  \\ \hline
Maximum long. acceleration ($m/s^2$) & 0.4 & 1.5  \\ \hline
Maximum long. deceleration ($m/s^2$) & 0.7 & 2.0  \\ \hline
Maximum lat. acceleration ($m/s^2$)  & 1.0 & 4.0  \\ \hline
\end{tabular}
\label{tab:trajectories}
\end{table}

\begin{figure}[htbp]
\centering
    \subfloat[Trajectory $T_1$]{
        \includegraphics[width=0.855\linewidth]{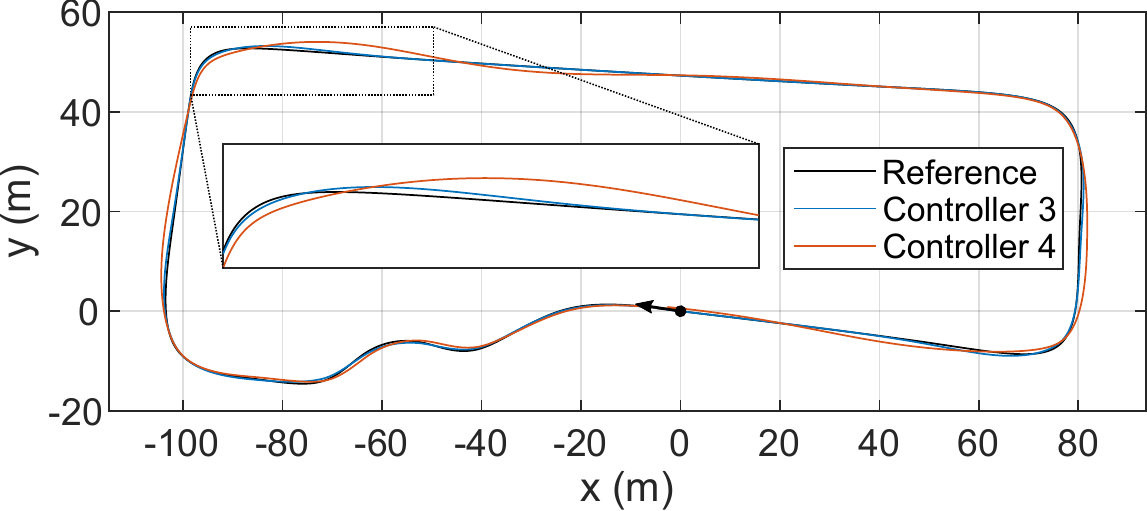}
        \label{fig:T1}}
        
    \subfloat[Trajectory $T_2$]{
        \includegraphics[width=0.9\linewidth]{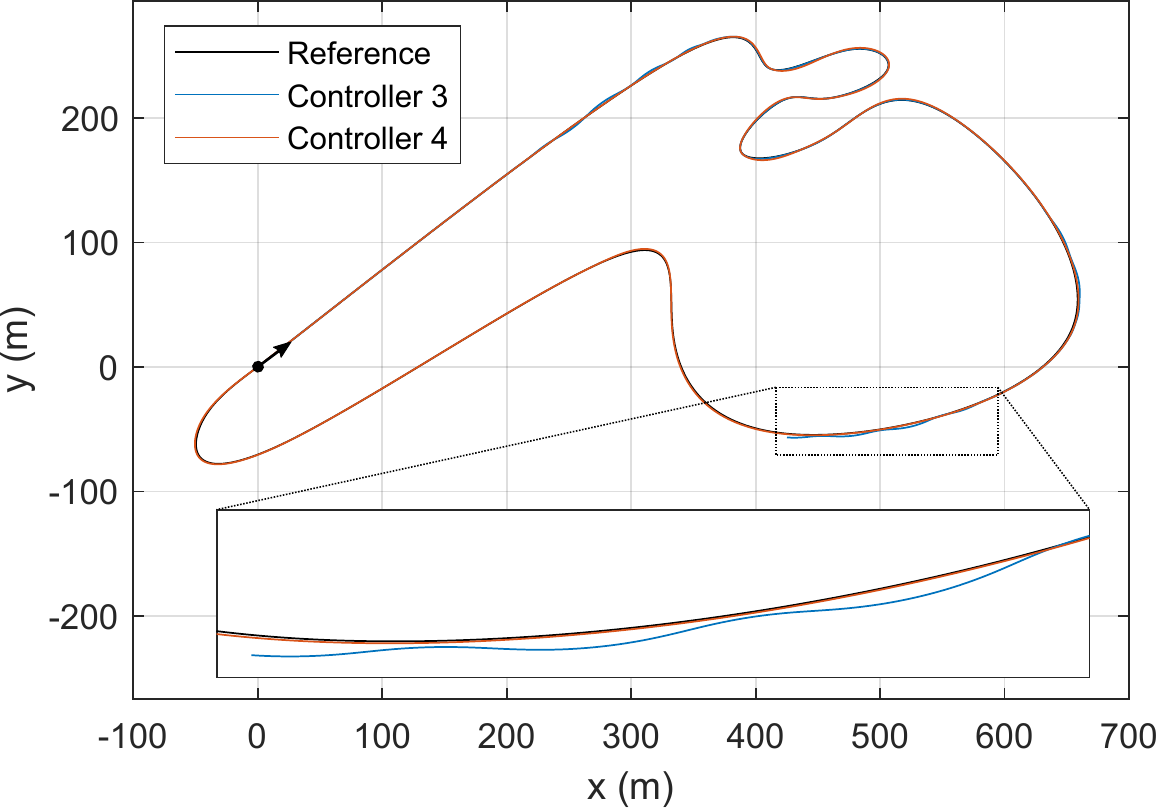}
        \label{fig:T2}}
        
    \subfloat[Longitudinal speed on $T_2$ for controller 3]{
        \includegraphics[width=0.85\linewidth]{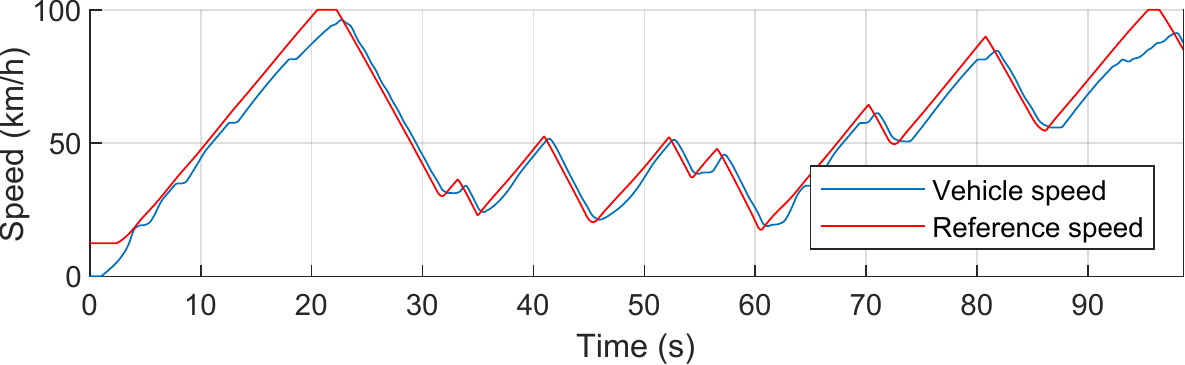}
        \label{fig:T2_speed}}
        
    \subfloat[Error metrics on $T_2$ for controller 3]{
        \includegraphics[width=0.85\linewidth]{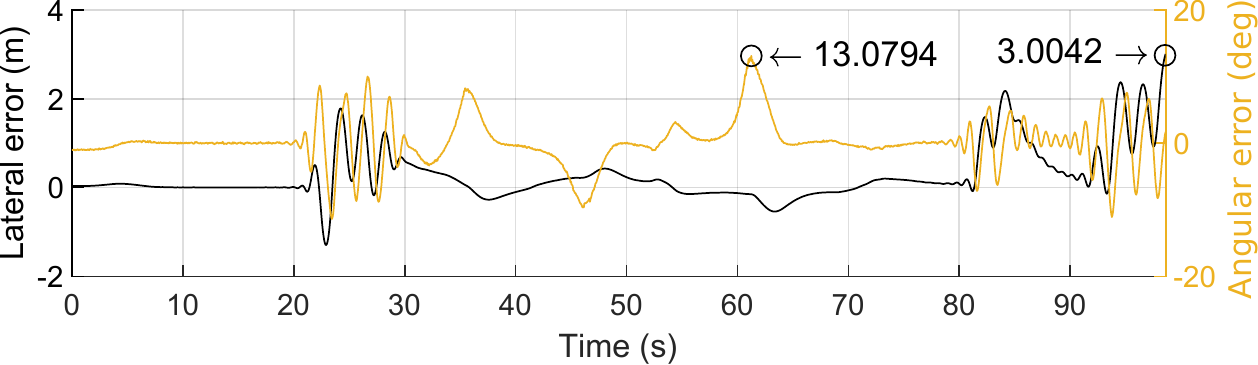}
        \label{fig:T2_error}}
\caption{Performance of controllers 3 and 4 on the benchmark trajectories}
\label{fig:Ts}
\end{figure}

Fig. \ref{fig:Ts} shows the performance of controllers 3 and 4 when applied on the extended model for test trajectories $T_1$ and $T_2$. As can be seen, controller 3 tracks better trajectory $T_1$, but becomes unstable when the speed increases in $T_2$ (simulation is stopped when the lateral error exceeds $3\, m$); however, although controller 4 has worse tracking performance in $T_1$, it remains stable even when the speed is high in $T_2$. This results show that controller 3 is less stable than controller 4, as it was specified by design.

\section*{Concluding remarks}

The aim of this work was to develop a design procedure to obtain the set of stabilizing MFC controllers and, among them, those that meet given frequency and time-response specifications.

The proposed algorithm, a non-linear adaptation of the Three Term Controller design procedure in \cite{bhattacharyya2018linear}, is able to preserve all the features of the original procedure, including frequency-response restrictions (gain and phase margins) and time-response restrictions (overshoot and settling time), giving the designer a powerful tool to design MFC controllers when a model of the system is available.

The proposed MFC tuning mechanism has been successfully applied to the design of lateral controllers for autonomous vehicles. Future work will address the design of MFC regulators for continuous-time systems without discretization.

\bibliography{ifacconf}             % bib file to produce the bibliography

\begin{thebibliography}{21}
\providecommand{\natexlab}[1]{#1}
\providecommand{\url}[1]{\texttt{#1}}
\providecommand{\urlprefix}{URL }
\expandafter\ifx\csname urlstyle\endcsname\relax
  \providecommand{\doi}[1]{doi:\discretionary{}{}{}#1}\else
  \providecommand{\doi}{doi:\discretionary{}{}{}\begingroup
  \urlstyle{rm}\Url}\fi

\bibitem[{Artu{\~n}edo et~al.(2019)Artu{\~n}edo, Godoy, and
  Villagra}]{artunedo2019decision}
Artu{\~n}edo, A., Godoy, J., and Villagra, J. (2019).
\newblock A decision-making architecture for automated driving without detailed
  prior maps.
\newblock In \emph{2019 IEEE Intelligent Vehicles Symposium (IV)}, 1645--1652.
  IEEE.

\bibitem[{Artu{ñ}edo et~al.(2021)Artu{ñ}edo, Villagra, and
  Godoy}]{Artunedo2021}
Artu{ñ}edo, A., Villagra, J., and Godoy, J. (2021).
\newblock Jerk-limited time-optimal speed planning for arbitrary paths.
\newblock \emph{IEEE Trans. on Intelligent Transport. Systems}, 1--15.

\bibitem[{Bhattacharyya et~al.(2018)Bhattacharyya, Datta, and
  Keel}]{bhattacharyya2018linear}
Bhattacharyya, S.P., Datta, A., and Keel, L.H. (2018).
\newblock \emph{Linear control theory: structure, robustness, and
  optimization}.
\newblock CRC press.

\bibitem[{d'Andr{\'e}a Novel et~al.(2010)d'Andr{\'e}a Novel, Fliess, Join,
  Mounier, and Steux}]{d2010mathematical}
d'Andr{\'e}a Novel, B., Fliess, M., Join, C., Mounier, H., and Steux, B.
  (2010).
\newblock A mathematical explanation via “intelligent” {PID} controllers of
  the strange ubiquity of {PID}s.
\newblock In \emph{18th Mediterranean Conference on Control and Automation,
  MED'10}, 395--400. IEEE.

\bibitem[{Fliess and Join(2013)}]{fliess.ijc13}
Fliess, M. and Join, C. (2013).
\newblock Model-free control.
\newblock \emph{International Journal of Control}, 86(12), 2228--2252.

\bibitem[{Fliess and Join(2021)}]{fliess2021alternative}
Fliess, M. and Join, C. (2021).
\newblock An alternative to proportional-integral and
  proportional-integral-derivative regulators: Intelligent
  proportional-derivative regulators.
\newblock \emph{Intl. J. of Robust and Nonlinear Ctrl.}, 1--13.

\bibitem[{Heged{\H{u}}s et~al.(2022)Heged{\H{u}}s, F{\'e}nyes, N{\'e}meth,
  Szab{\'o}, and G{\'a}sp{\'a}r}]{hegedHus2022design}
Heged{\H{u}}s, T., F{\'e}nyes, D., N{\'e}meth, B., Szab{\'o}, Z., and
  G{\'a}sp{\'a}r, P. (2022).
\newblock Design of model free control with tuning method on ultra-local model
  for lateral vehicle control purposes.
\newblock In \emph{2022 American Control Conf.}, 4101--4106. IEEE.

\bibitem[{Jiang and Astolfi(2018)}]{jiang2018lateral}
Jiang, J. and Astolfi, A. (2018).
\newblock Lateral control of an autonomous vehicle.
\newblock \emph{IEEE Transactions on Intelligent Vehicles}, 3(2), 228--237.

\bibitem[{Keel and Bhattacharyya(2002)}]{keel2002root}
Keel, L. and Bhattacharyya, S. (2002).
\newblock Root counting, phase unwrapping, stability and stabilization of
  discrete time systems.
\newblock \emph{Linear Algebra Appl.}, 351, 501--518.

\bibitem[{Lee et~al.(2018)Lee, Kim, and Kim}]{7886365}
Lee, D., Kim, K.S., and Kim, S. (2018).
\newblock Controller design of an electric power steering system.
\newblock \emph{IEEE Trans. on Ctrl. Systems Tech.}, 26(2), 748--755.

\bibitem[{Li et~al.(2022)Li, Yuan, Li, and Zhu}]{li2022revisit}
Li, W., Yuan, H., Li, S., and Zhu, J. (2022).
\newblock A revisit to model-free control.
\newblock \emph{IEEE Transactions on Power Electronics}, 37(12), 14408--14421.

\bibitem[{Moreno-Gonzalez et~al.(2022)Moreno-Gonzalez, Artu{\~n}edo, Villagra,
  Join, and Fliess}]{moreno2022speed}
Moreno-Gonzalez, M., Artu{\~n}edo, A., Villagra, J., Join, C., and Fliess, M.
  (2022).
\newblock Speed-adaptive model-free lateral control for automated cars.
\newblock In \emph{Joint 8th IFAC Symposium on System Structure and Control,
  17th IFAC Workshop on Time Delay Systems, 5th IFAC Workshop on Linear
  Parameter Varying Systems, IFAC 2022}.

\bibitem[{Pacejka and Bakker(1992)}]{Pacejka1992}
Pacejka, H. and Bakker, E. (1992).
\newblock The magic formula tyre model.
\newblock \emph{Vehicle System Dynamics}, 21, 1--18.

\bibitem[{Rajamani(2011)}]{rajamani2011vehicle}
Rajamani, R. (2011).
\newblock \emph{Vehicle dynamics and control}.
\newblock Springer Science \& Business Media.

\bibitem[{Sanyal(2022)}]{sanyal2022discrete}
Sanyal, A. (2022).
\newblock Discrete-time data-driven control with h{\"o}lder-continuous
  real-time learning.
\newblock \emph{International Journal of Control}, 95(8), 2175--2187.

\bibitem[{Villagra(2023)}]{villagra2023interplay}
Villagra, J. (2023).
\newblock Interplay between decision and control.
\newblock In \emph{Decision-Making Techniques for Autonomous Vehicles},
  193--213. Elsevier.

\bibitem[{Villagra et~al.(2009)Villagra, d'Andr{\'e}a Novel, Choi, Fliess, and
  Mounier}]{villagra.ijvas09}
Villagra, J., d'Andr{\'e}a Novel, B., Choi, S., Fliess, M., and Mounier, H.
  (2009).
\newblock Robust stop-and-go control strategy: an algebraic approach for
  non-linear estimation and control.
\newblock \emph{Intl. J. of Vehicle Auton. Syst.}, 7(3-4), 270--291.

\bibitem[{Villagra et~al.(2020)Villagra, Join, Haber, and
  Fliess}]{villagra2020model}
Villagra, J., Join, C., Haber, R., and Fliess, M. (2020).
\newblock Model-free control for machine tools.
\newblock In \emph{21st IFAC World Congress, IFAC 2020}.

\bibitem[{Wang et~al.(2020)Wang, Xu, Tian, and Tang}]{wang2020alpha}
Wang, H., Xu, H., Tian, Y., and Tang, H. (2020).
\newblock $\alpha$-variable adaptive model free control of irehave upper-limb
  exoskeleton.
\newblock \emph{Adv. in Eng. Software}, 148, 102872.

\bibitem[{Yahagi and Kajiwara(2022)}]{Yahagi2022}
Yahagi, S. and Kajiwara, I. (2022).
\newblock Non-iterative data-driven tuning of model-free control based on an
  ultra-local model.
\newblock \emph{IEEE Access}, 10, 72773--72784.

\bibitem[{Zainal et~al.(2017)Zainal, Rahiman, and Baharom}]{zainal2017yaw}
Zainal, Z., Rahiman, W., and Baharom, M. (2017).
\newblock Yaw rate and sideslip control using {PID} controller for double lane
  changing.
\newblock \emph{J. Telec. Electr. Comp. Eng.}, 9, 99--103.

\end{thebibliography}
                                                     % with bibtex (preferred)
                                                     
\end{document}